\documentclass{aa}
\input epsf
\def\spose#1{\hbox to 0pt{#1\hss}}
\def\mdot{\spose{\raise 7.0pt\hbox{\hskip 5.0pt{\char '056}}}M}

\begin{document}
\thesaurus{08
      (08.06.3;  
       08.09.2 WR124;  
       08.23.2;  
       09.08.1;  
       09.09.1 M1-67)} 

\title{Wolf-Rayet nebulae as tracers of stellar ionizing fluxes: I. M1-67}
\author{Paul A. Crowther \inst{1}, A. Pasquali \inst{2}, Orsola
De Marco \inst{1,3}, W. Schmutz \inst{3,4}, D.J. Hillier \inst{5},
A. de Koter \inst{6} }

\offprints{P.A. Crowther (pac@star.ucl.ac.uk)}
\institute{Department of Physics and Astronomy, University College London,
           Gower Street, London WC1E 6BT, U.K.
           \and
           ST-ECF/ESO, Karl-Schwarzschild-Strasse 2, D-85748, Garching
           bei M\"unchen, Germany 
           \and
           Institut f\"ur Astronomie, ETH-Zentrum, CH-8092 Z\"urich,
           Switzerland
           \and
           Physikalisch-Meteorologisches Observatorium Davos, CH-7260 
           Davos Dorf, Switzerland
           \and
           Department of Physics and Astronomy, University of Pittsburgh,
           3941 O'Hara Street, Pittsburgh, PA 15260, USA 
           \and
           Astronomical Institute `Anton Pannekoek', University of 
           Amsterdam, Kruislaan 403, 1098 SJ Amsterdam, The Netherlands}
\date{today}
\authorrunning{Crowther et al.}
\titlerunning{Wolf-Rayet nebulae I. M1-67}
\maketitle
\begin{abstract}
We use WR124 (WN8h) and its associated nebula M1--67, to test 
theoretical non-LTE models for Wolf-Rayet (WR) stars. Lyman continuum 
ionizing flux distributions derived from a stellar analysis of WR124,
are compared with nebular properties via photo-ionization modelling.
Our study demonstrates the significant role that line blanketing plays
in affecting the Lyman ionizing energy distribution of WR stars, of
particular relevance to the study of H\,{\sc ii} regions containing young stellar
populations.

We confirm previous results 
that non-line blanketed
WR energy distributions fail to explain the observed nebular 
properties of M1--67, such that the predicted ionizing spectrum is too hard.
A line blanketed analysis of WR124 is carried out using the method
of Hillier \& Miller (1998), with stellar properties in accord with 
previous results, except that the inclusion of clumping in the
stellar wind reduces its 
wind performance factor to only $\sim$2. The ionizing spectrum of the
line blanketed model is much softer than for a comparable temperature 
unblanketed case, such that negligible flux is emitted with energy above
the He\,{\sc i} $\lambda$504 edge. Photo-ionization modelling, incorporating 
the observed radial density distribution for M1--67 
reveals excellent agreement with the observed nebular electron 
temperature, ionization balance and line strengths.
An alternative stellar model of WR124 is calculated, following the technique 
of de Koter et al. (1997), augmented to include line blanketing following 
Schmutz et al. (1991). Good consistency is reached regarding the stellar
properties of WR124, but agreement with the nebular properties of M1--67
is somewhat poorer than for the Hillier \& Miller code.


\end{abstract}

\section{Introduction}

The Lyman ionizing energy distributions of hot, massive stars are
important in the study of young starburst regions and galaxies, via
population synthesis codes (e.g. Leitherer et al. 1999). 
However, the interstellar medium (ISM) conspires to prevent this energy from
reaching the Earth's atmosphere.
Interstellar dust severely degrades transmitted fluxes at
ultraviolet and optical wavelengths, while
the Lyman continuum of hot stars is completely
absorbed by intervening hydrogen atoms for all but a few nearby B stars
(e.g. Cassinelli et al. 1995).
Therefore, indirect measurements of the extreme UV flux are necessary.

Until recently, the standard method of obtaining the ionizing 
distributions of hot stars was to rely on line blanketed,
plane parallel, LTE model predictions from Kurucz (1991). However, 
the need to consider non-LTE effects, spherical geometry and
line blanketing effects has recently led to the development of 
complex model atmospheres (e.g. Hubeny \& Lanz 1995; 
de Koter et al. 1997; Pauldrach et al. 1998), with 
Lyman continuum ionizing flux distributions obtained 
via synthesis of the accessible UV and optical stellar spectrum. 
However, these different codes predict quite different ionizing spectra 
for early-type stars of identical temperatures. While attempts at 
testing theoretical stellar O-type models via
observations of stellar clusters have been made (Stasi\'{n}ska \&
Schaerer 1998; Oey \& Kennicutt 1998), definitive results require
studies of {\it individual} stars and their associated H\,{\sc ii} regions.

In the case of Wolf-Rayet (WR) stars,
evolutionary synthesis models for young starbursts known to contain
these stars, known as `WR galaxies', still rely on unblanketed,
non-LTE model atmospheres of Schmutz et al. (1992).
What effect does line blanketing have on the predicted Lyman continua
of WR stars? Recent theoretical calculations by Crowther (1999) suggest
that the effect can be substantial, such that unblanketed WR models
overestimate the hardness of their ionizing spectra, which is supported
by observations of extra-galactic giant H\,{\sc ii} regions by
Bresolin et al. (1999).

\begin{table}
\begin{center}
\caption{Summary of basic stellar and nebular properties of
WR124/M1--67.}
\label{table1}
\begin{tabular}{l@{\hspace{0.5mm}}l@{\hspace{0.5mm}}l}
\hline
Quantity & Measurement & Reference \\
\hline
$m_{v}$      & 11.58 mag & Massey (1984) \\
$(b-v)$      & 0.81 mag & Massey (1984) \\
$m_{\rm K}$  &  7.71 mag & van der Hucht et al. (1985) \\
$E_{\rm B-V}$& 1.18 mag & Crowther et al. (1995b) \\
Distance     & 4--5 kpc & Crawford \& Barlow (1991) \\
Sp. Type     & WN8h    & Smith et al. (1996) \\
\noalign{\smallskip}
R   & 55--60 arcsec    & Grosdidier et al. (1998) \\
$N_{e}$, $T_{e}$  & 1050 cm$^{-3}$, 6200K  & Esteban et al. (1991) \\
$F_{{\rm H}\alpha}$& 2.84$\times$10$^{-11}$ erg\,cm$^{-2}$s$^{-1}$ & Grosdidier et al. (1998) \\
\hline
\end{tabular}
\end{center}
\end{table}

In principle, H\,{\sc ii} regions associated with their
central star are ideal tracers of the Lyman ionizing flux distributions.
However, suitable nebulae are rare, since their 
properties are often
poorly known. The principal studies for WR stars are those of
Rosa \& Mathis (1990) and Esteban et al. (1993) who
combined WR model fluxes from Schmutz et al. (1989, 1992) with
observed properties of WR ring nebulae, to investigate the properties 
of the central stars. Esteban et al. varied the temperatures of unblanketed
WR models until agreement was reached between the observed nebular
properties and those predicted by photo-ionization modelling. 
In general, comparisons with (independent) stellar analyses of the
central stars was found to be reasonable, except that lower temperatures were
required from the photo-ionization models for late-type WN (WNL) stars, 
especially the Galactic WN8 star WR124 and its associated 
H\,{\sc ii} region, M1--67, for which no agreement was achieved.

In this work, we shall depart from the technique of Esteban et al. in that
stellar properties of WR124 are fixed from a spectroscopic study, which are
then tested against the nebular properties of M1--67 via photo-ionization
modelling using {\sc cloudy} (Ferland 1996). Sirianni et al. (1998)
showed M1--67 to be very clumpy. Hubble Space Telescope (HST) imaging of
M1--67 by Grosdidier et al. (1998) revealed an astonishingly complex nebula,
allowing
measurements of the nebular radial density distribution and integrated
H$\alpha$ flux.  Our work therefore represents the first attempt
to  test the reliability of predicted WR Lyman continuum distributions
using both unblanketed (Hillier 1987) and
blanketed models (following Hillier \& Miller 1998, 1999) plus robust nebular
properties. In our study we compare these predictions with the code
of de Koter et al. (1997), augmented to allow for line blanketing
following Schmutz (1994, 1997).
In this way, we will be able to quantitatively assess the effects of
different codes and line blanketing techniques on the ionizing 
energy produced by WR stars.

In Sect. 2 we will discuss observations of WR124 and M1--67.
In Sect. 3 the model atmospheres codes are introduced and discussed,
with stellar results compared in Sect. 4. The photo-ionization modelling
technique is discussed and results are presented in Sect. 5. Finally,
     conclusions are drawn in Sect. 6.

\section{Observations of WR124 and its associated nebula, M1--67}

We discuss  the observed properties of WR124 and M1--67 in this
section, and provide a summary in Table~\ref{table1}.
Unlike most massive WR nebulae, which are difficult
to identify, M1--67 was so striking that it was included in the 
Bertola (1964) Planetary Nebula catalogue (PK\,50+3$^{\circ}$ 1) with
WR124 as its central star. Cohen \& Barlow (1975) instead
proposed a massive WN origin, which was supported by
Solf \& Carsenty (1982) and Esteban et al. (1991), who
found nebular abundances to be consistent with processed stellar ejecta,
indicative of a massive central star. Confirmation of a non-PN origin
was made by Crawford \& Barlow (1991)  who estimated a  distance of
4--5 kpc from interstellar Na\,{\sc i} D$_2$ observations. (In order for
M1-67 to be a PN, its distance should be $\sim$460\,pc according
to van der Hucht et al. 1985).

\begin{figure}
\epsfxsize=8.8cm \epsfbox[70 150 430 680]{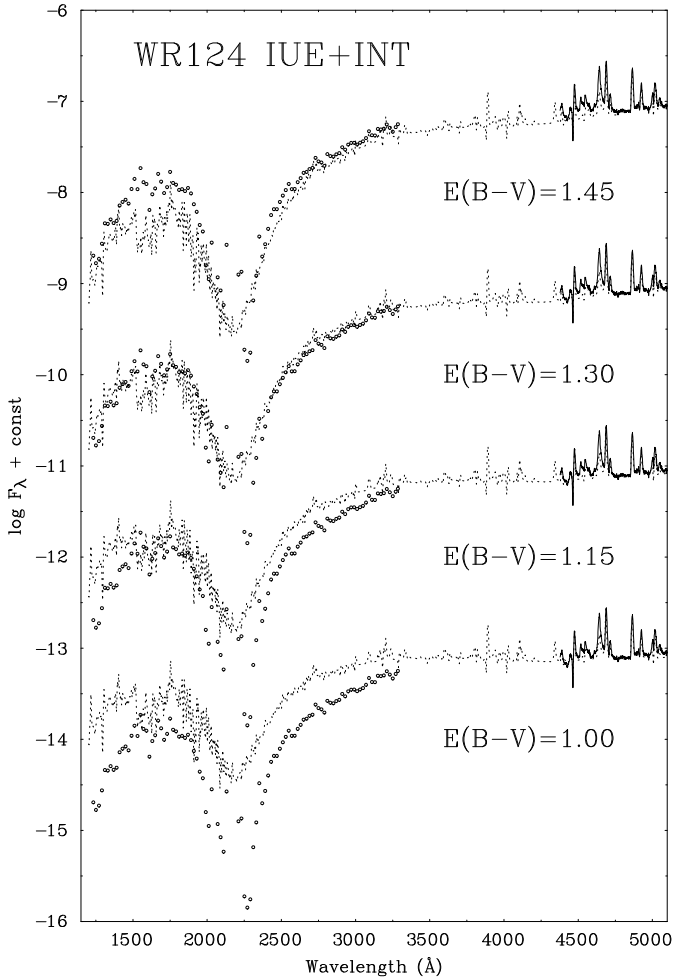}
\caption{Comparison between the spectrophotometry of WR124 (open circles:
IUE; solid line: INT) and theoretical {\sc cmfgen} models, reddened by
$E_{\rm B-V}$=1.0, 1.15, 1.3 and 1.45 mag (dotted lines) according to
Cardelli et al. (1989). For clarity, comparisons are offset vertically
by 0, +2, +4 and +6 units.}
\label{fig1}
\end{figure}

\subsection{WR124}\label{reddening}

WR124 (alias 209\,BAC, He~2--427) is a Galactic, heavily reddened WN8h star
that is well known as the ionizing star of M1--67, with a high heliocentric
recession velocity of $\sim$200 km\,s$^{-1}$ (Merrill 1936).
Crowther et al. (1995b) presented optical spectroscopy of WR124, including
comparisons with other late-type WN  (WNL) stars.

For the current analysis, spectrophotometry obtained in 1991
September at the 2.5m Isaac Newton Telescope (INT),
covering $\lambda\lambda$4400--7300 (resolution 2--3\AA)
are used, taken from Crowther et al. (1995b). Observations
obtained at the Deutsch--Spanisches Astronomisches Zentrum (DSAZ) 2.2m
observatory in June 1991 provide an additional blue dataset covering
$\lambda\lambda$3400--4400 ($\sim$2\AA\ resolution), details about which
can be obtained from Hamann et al. (1995). 

New near-IR spectroscopy of WR124 was obtained in 1998 July at the
3.8m U.K. Infrared Telescope (UKIRT) with the cooled grating spectrograph 
CGS4, the long (300mm) camera, a 256$\times$256 InSb array and a 
40\,l/mm grating. First order observations provided data 
covering 1.66--2.30$\mu$m (R$\sim$800) and 2.98--3.62$\mu$m 
(R$\sim$1300) with a slit width of 0.6$''$ (1 pixel). This data set 
was bias-corrected, flat-fielded, extracted and sky-subtracted using 
{\sc cgs4dr} (Daly \& Beard 1992). Subsequent reductions and analysis
were carried out using {\sc figaro} (Meyerdierks 1993) and {\sc dipso}
(Howarth et al. 1995). In order to remove atmospheric features, 
the observations were divided by a standard star (whose spectral features 
were artificially removed) observed at around the same time and similar 
air mass. In regions of low atmospheric transmission at UKIRT the 
reliability of line shape and strength must be treated with caution 
(e.g. Paschen $\alpha$).

The reddening to M1--67 is extremely high, with previous estimates of
$E_{\rm B-V}$ in the range 0.90 (Esteban et al. 1991) to 1.50
(Solf \& Carsenty 1982). Analysis of the stellar spectrum by 
Crowther et al. (1995b) indicated $E_{\rm B-V}$=1.18 mag, which implied a distance
of $\sim$6kpc, assuming a typical $M_{v}$=$-6.0$ mag for WN8 stars.
Here we attempt an improved stellar reddening determination by
including International Ultraviolet Explorer (IUE)
newly extracted spectra (INES, Rodriguez-Pascual et al. 1998) of
low resolution (LORES) datasets. The quality of LORES INES datasets are 
preferred to NEWSIPS (Schartel \& Skillen 1998).
Four large aperture IUE datasets of WR124 were obtained between
1978--1986, including 3 long wavelength (LWR) datasets, totalling
270 minutes exposure time, plus 1  short wavelength (SWP) dataset
of 212 minutes. 

Pre-empting the results from the next section, we compare
spectrophotometry with reddened, line-blanketed model atmosphere distributions
obtained with the {\sc cmfgen} code (see Sect.~\ref{cmfgen}) in 
Fig.~\ref{fig1},
 indicating 
$E_{\rm B-V}$$\sim$1.3 mag ($A_{\rm V}$=4.0), 0.1--0.2 mag higher than previous results based 
solely on optical stellar datasets. A comparison with alternative {\sc
isa}-wind 
(see Sect.~\ref{ISA}) model fluxes indicates an equivalent reddening.

Our measurement is in reasonable agreement with the nebular
study of M1--67 by Chu \& Treffers (1981) who obtained $E_{\rm B-V}$=1.23
from a comparison of its 1.4\,GHz and H$\alpha$ flux.  In contrast, Esteban 
et al. (1991) obtained an significantly lower value of
$E_{\rm B-V}$=0.90 mag from nebular Balmer line strengths.

Since the reddening towards WR124 is so high, we prefer to select
the distance from the K-band absolute magnitude, $M_{\rm K}$, where 
the interstellar reddening is much lower. Using WN8--9 stars at 
known distance in our Galaxy or the Large Magellanic Cloud (LMC) as calibrators
(see Morris et al. 1999 and references therein) 
we obtain $M_{\rm K}$=$-$6.2$\pm$0.3 mag. 
Using the observed K-band magnitude of WR124 from van der Hucht et al. 
(1985) plus $A_{\rm K}$ = 0.12 $\times A_{\rm V}$ = 0.49 
(Cardelli et al. 1989) we
find a distance modulus of 13.44 (i.e. 4.9 kpc), in accord with
Crawford \& Barlow (1991). Consequently, we adopt a distance of 5 
kpc to WR124, i.e. $M_{\rm K}$=$-$6.25 mag. Note that the resulting 
V-band absolute magnitude of $M_{\rm V}$=$-$6.1 mag is in excellent 
agreement with LMC WN8--9 stars (Crowther \& Smith 1997).

\subsection{M1--67}

M1--67, discovered by Minkowski (1946), has a clumpy, ejecta-type,
morphology (Chu 1981), with an
angular diameter of 90--120 arcsec (Solf \& Carsenty 1982; Grosdidier
et al. 1998).
High spatial resolution spectroscopy of M1--67 was obtained by Esteban et al.
(1991), revealing nebular conditions of $N_{e}$$\sim$10$^{3}$ and
$T_{e}$$\sim$6,200K. Relative to normal H\,{\sc ii} regions, oxygen is
deficient, with nitrogen enhanced, such that N/O ($\approx$3). Their results
suggest that O has been processed into N mainly through the ON cycle, and
confirm that the M1--67 nebula is composed of stellar ejectra. Esteban 
et al. estimated
an upper limit of 0.2 for the ratio between the ISM mass swept-up by
M1--67 and the nebular mass, suggesting that relatively little mixing 
with the local interstellar gas has occurred.

Fabry-Perot interferometry (Chu \& Treffers 1981) and long-slit
spectroscopy (Sirianni et al. 1998) indicate a complicated velocity
field, with two different motions: a spherical, hollow shell,
expanding at 46 km\,s$^{-1}$, plus a `bipolar' outflow with a larger velocity
of 88 km\,s$^{-1}$.
Sirianni et al. (1998) interpreted this structure as the result of two
subsequent outbursts. Grosdidier et al. (1998) used the very high spatial
quality of the Wide Field and Planetary Camera 2 (WFPC2) aboard HST to
reveal filamentary, chaotic substructures, with a number of `bullets'
in the nebula, which seem to concentrate within two conical regions
in agreement with the previous results of Sirianni et al. (1998).
Globally, Grosdidier et al. (1998) found that the de-projected radial
density distribution of M1--67 is well described by a power law,
$N_e (r) \sim r^{-0.8}$, with a cutoff at $r \ge$50$''$.

\section{Stellar atmosphere codes}

In this section we introduce and utilise the codes that are used to
carry out spectral  synthesis of WR124. {\sc cmfgen}
(Hillier 1987, 1990; Hillier \& Miller 1998, 1999) solves the
transfer equation in the co-moving frame, subject to statistical
and radiative equilibrium, assuming an expanding, spherically-symmetric,
homogeneous or clumped, 
atmosphere. Populations and ionization
structure are consistent with the radiation field. 

To take into account the (microscopic and macroscopic) velocity 
structure expected in a real WR star, a uniform Doppler line width 
of $V_{\rm Dop}$=50 km\,s$^{-1}$ is assumed.
Since this is substantially greater than the (species dependent) thermal 
Doppler width, we refer to it as a `broad-line' approximation. 
All lines within some multiple of $V_{\rm Dop}$ can interact within 
the Sobolev resonance zone.
Because of the large line width used, the interaction between two 
lines may be overestimated in the broad-line approximation.

A simplifying `super level' approach is used for individual levels 
(Anderson 1989), particularly for iron-group elements.
In this approach, several levels of similar energies and properties
are treated as a single `super level', with only the populations of
the super level included in the solution of the rate equations. The
population of an individual atomic level in the full model atom is
determined by assuming that it has the same departure coefficient as 
the corresponding super level to which it belongs.

Unfortunately, individual line blanketed co-moving frame
calculations are generally computationally demanding, despite
the use of super levels, so we first need to confirm that individual
case studies, such as this, are in accord with observation.
A computationally quick method is to (i) solve the transfer problem 
in the Sobolev approximation rather than the co-moving frame, which the
code {\sc isa}-wind does (de Koter et al. 1993, 1997),
and (ii) consider
line blanketing via Monte Carlo sampling following Schmutz
(1994, 1997) allowing the opacity of a huge number of lines to be
considered.  How consistent are results obtained with these different
methods?

\begin{table*}
\caption[]{Summary of model atom used in {\sc cmfgen} and {\sc isa}-wind
radiative transfer calculations, including full levels ($N_{\rm F}$), 
super levels ($N_{\rm S}$) 
and total number of transitions ($N_{\rm Trans}$), following 
Hillier \& Miller (1998) and de Koter et al. (1997). Note that transitions
considered in {\sc cmfgen} include those between individual terms in 
multiplets. The Monte Carlo calculation used to derive line blanketing 
accounts for transitions of most metals, including Al, Si and Fe.}
\label{table2}
\begin{flushleft}
\begin{tabular}{lrrrrrll}
\hline\noalign{\smallskip}
Ion & \multicolumn{1}{c}{$N_{\rm S}$} & \multicolumn{2}{c}{$N_{\rm F}$} 
& \multicolumn{2}{c}{$N_{\rm Trans}$} & \multicolumn{2}{c}{Details} \\
   & {\sc cmfgen} & {\sc cmfgen} & {\sc isa}-wind & 
     {\sc cmfgen} & {\sc isa}-wind & {\sc cmfgen} & {\sc isa}-wind \\
\hline\noalign{\smallskip}
H\,{\sc i}    & 10 & 30 & 10 & 435 & 45 & $n \le$30. & $n\le$10 \\
H\,{\sc ii}   &  1 &  1 &  1 &     &    & \\
He\,{\sc i}   & 29 & 51 & 51 & 579 & 691& $n \le$20. & $n\le$20 \\ 
He\,{\sc ii}  & 13 & 30 & 20 & 435 &190 & $n \le$30. & $n\le$20 \\ 
He\,{\sc iii} &  1 &  1 &  1 &     &    & \\
C\,{\sc iii}  & 21 & 38 & -- & 147 & -- & $nl \le$2s4f$^{1}$F$^{\circ}$&-- \\
C\,{\sc iv}   &  9 & 14 & 12 & 48  & 66 & $n \le 4$. & $n\le$7 \\ 
C\,{\sc v}    &  1 &  1 &  1 &     &    & \\
N\,{\sc ii}   & 23 & 41 &  3 & 144 &  3 & $nl \le$2p3d$^{1}$P$^{\circ}$ & $nl \le$2p$^{2}$ $^{1}$S \\ 
N\,{\sc iii}  & 33 & 48 & 40 & 407 & 561& $nl \le$2p4p$^{2}$D & $nl \le$2p4p$^{2}$P          \\ 
N\,{\sc iv}   & 22 & 38 & 34 & 144 & 561& $nl \le$2p3d$^{1}$D$^{\circ}$ & $nl \le$2s4f$^{1}$F$^{\circ}$ \\ 
N\,{\sc v}    &  9 & 15 & 26 &  42 & 325& $n \le 5$ & $n \le 10$\\ 
N\,{\sc vi}   &  1 &  1 &  1 &     &    & \\
Al\,{\sc iii} & 17 & 45 & -- & 362 & -- & $n \le 10$ & --\\
Al\,{\sc iv}  &  1 &  1 & -- &     &    & \\
Si\,{\sc iii} & 25 & 45 & -- & 170 & -- & $nl \le$3s5f$^{1}$F$^{\circ}$ &--\\
Si\,{\sc iv}  & 17 & 28 & -- & 129 & -- & $n \le 6$ & -- \\
Si\,{\sc v}   &  1 &  1 & -- &     &    & \\
Fe\,{\sc iii} & 52 &242 & -- &1840 & -- & $nl \le$3d$^5$($^{4}$F)5s$^{5}$F&-- \\
Fe\,{\sc iv}  & 21 &280 & -- &3538 & -- & $nl \le$3d$^4$($^{1}$G)4p$^{2}$P$^{\circ}$&-- \\
Fe\,{\sc v}   & 19 &182 & -- &1803 & -- & $nl \le$3d$^3$($^{2}$D)4p$^{1}$P$^{\circ}$&-- \\
Fe\,{\sc vi}  & 10 &80  & -- & 722 & -- & $nl \le$3d$^2$($^{1}$S)4p$^{2}$P$^{\circ}$&-- \\
Fe\,{\sc vii} &  1 &  1 & -- &     &    & \\
\noalign{\smallskip}\hline
              &337 &1214&181&10945& 2442 & \\
\noalign{\smallskip}\hline
\end{tabular}
\end{flushleft}
\end{table*}

\subsection{CMFGEN}\label{cmfgen}

The model calculations are based on the iterative technique
of Hillier (1987, 1990) named {\sc cmfgen}.
Allowance is made for line blanketing  and clumping following the
formulation of Hillier \& Miller (1998) to which the reader is referred
for specific details. The model atom contains
hydrogen, helium, carbon, nitrogen, silicon and iron as shown in
Table~\ref{table2}. 
Details of each ion are included, such that for
hydrogen, $n$=1 to 30 full levels are considered, which are grouped into 
10 super levels. 
Test calculations indicated that $\sim$5000 iron transitions with 
$gf \le 10^{-4}$ have negligible influence on the emergent spectrum, but 
required considerable computational effort, so these have been excluded. 
Within a super-level, representing a group 
of full levels, populations are assumed to be in LTE. 
Elemental abundances other than hydrogen and helium are fixed at
cosmic (Si=0.1\% by mass, Fe=0.2\% by mass) or approximately
CNO-equilibrium (C=0.03\%, N=1 to 2\% by mass) values for a 
solar metallicity environment. 

The stellar radius ($R_{\ast}$) is defined
as the inner boundary of the model atmosphere and is located at
Rosseland optical depth of $\sim$20 with the stellar temperature ($T_{\ast}$)
defined by the usual Stefan-Boltzmann relation. Similarly, the effective
temperature ($T_{\rm eff}$) relates to the radius ($R_{\rm 2/3}$)
at which the Rosseland optical depth equals 2/3.

\subsection{ISA-wind}\label{ISA}

The improved Sobolev approximation code ({\sc isa}-wind) is described in
detail by de Koter et al. (1993, 1997). The principal differences with 
{\sc cmfgen} relate to: (a) the treatment of the line radiation transfer; 
(b) the wind electron temperature, which assumes a grey, LTE atmosphere 
rather than radiative equilibrium, where the temperature is fixed at 
a pre-assigned value in the outer regions, 
 such that lines originating in the outer wind 
(e.g. He\,{\sc i} $\lambda$10830) may be incorrectly predicted; 
(c) the velocity structure deep in the atmosphere; 
(d) the specific atomic model
treated, as listed in Table~\ref{table2}; (e) the neglect of clumping.

\begin{figure*}
\epsfxsize=15cm \epsfbox[25 240 525 780]{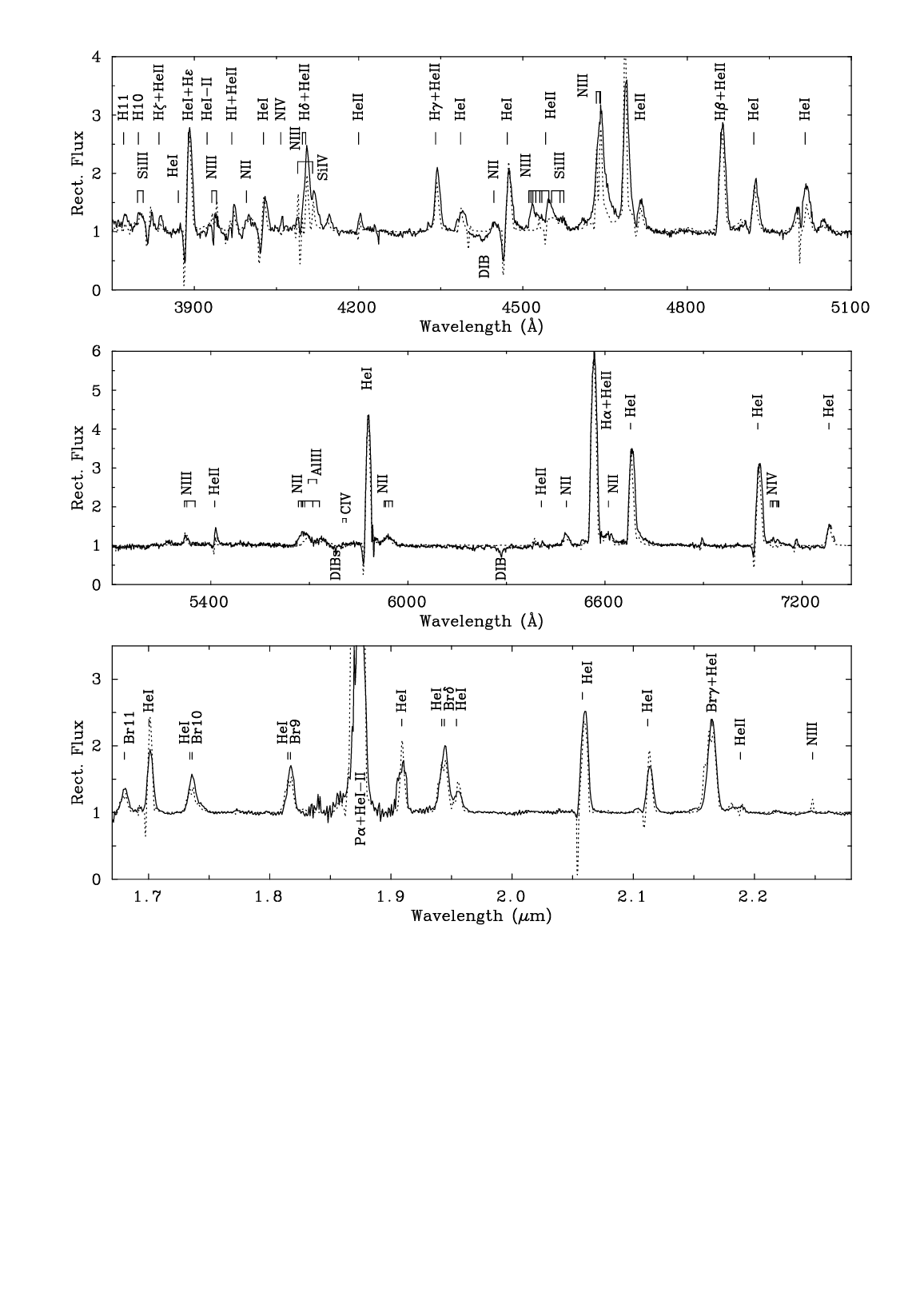}
\caption{Comparison between the rectified optical (INT) and near-IR (UKIRT)
spectrum of WR124 (solid lines) with the synthetic spectrum 
produced by {\sc cmfgen}, accounting for line blanketing and clumping.}
\label{fig2}
\end{figure*}

\begin{figure*}
\epsfxsize=15cm \epsfbox[25 590 525 770]{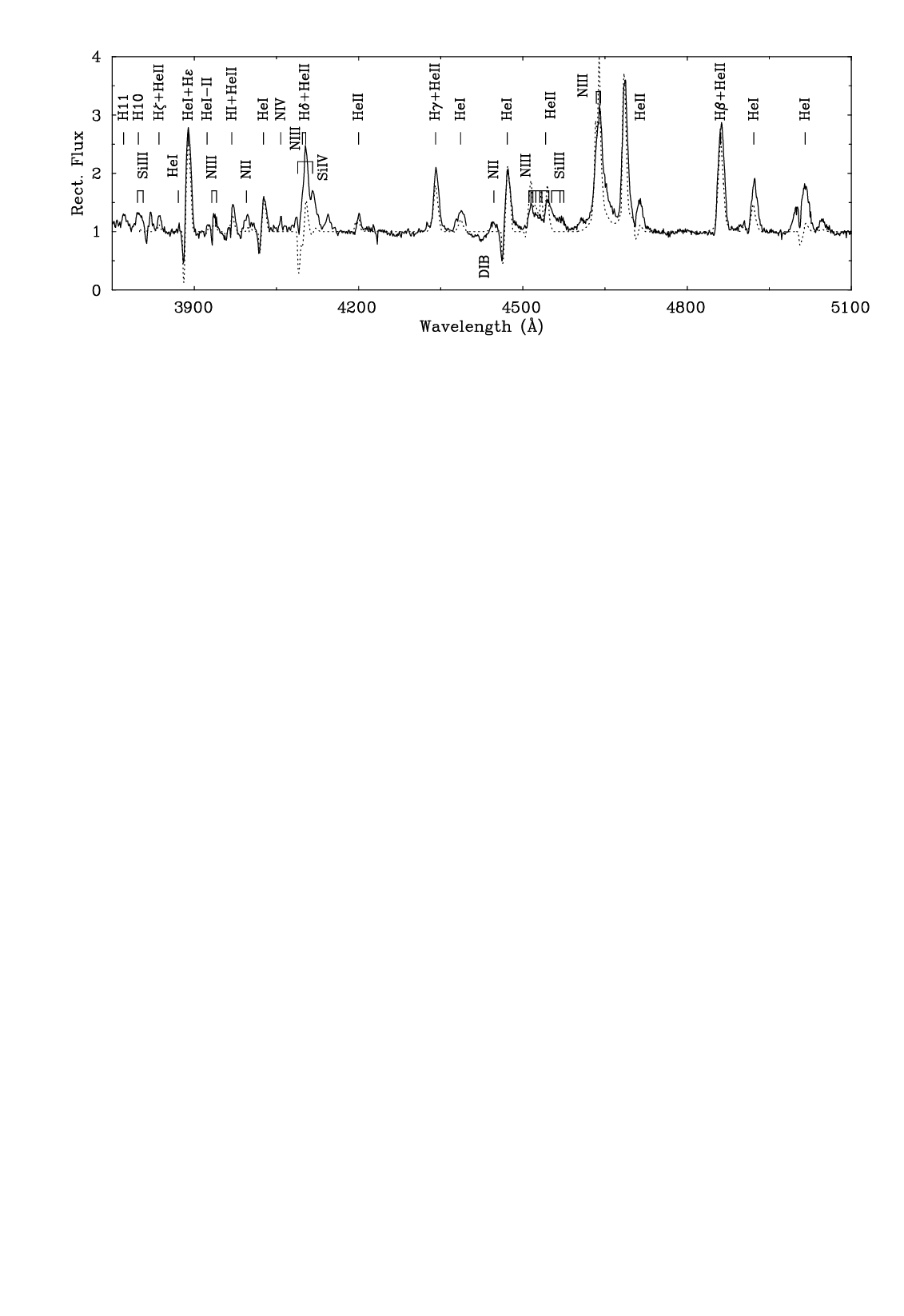}
\caption{Comparison between the rectified optical (INT) 
spectrum of WR124 (solid lines) with the synthetic spectrum 
produced by {\sc isa}-wind, accounting for line blanketing.
Spectral comparisons in the red and near-IR are comparable 
to {\sc cmfgen}, except for those lines discusssed in the text.}
\label{fig2a}
\end{figure*}

{\sc isa}-wind solves the line radiation transfer using an improved version
of the Sobolev approximation. In co-moving frame codes, line photons with
a finite (given) width interact with the continuum over a range of
depth and frequency points, while in the Sobolev approximation the continuum
opacity and source function are considered constant within the line
resonance volume. In the improved version of the Sobolev code, 
{\sc isa}-wind, absorption
of line radiation by the continuum is taken into account within the
resonance volume. The Sobolev approximation 
should be ideal for the large velocity gradients of WR outflows.
It introduces great simplification 
in the rate equations, and 
makes the overall iteration process about 10 times faster 
than a co-moving frame code, for the same number or 
ionic levels and grid points.

Turning to line blanketing, an iterative
technique including the Monte Carlo method of
Schmutz et al. (1991) and Schmutz (1994, 1997) is used.
The method allows the computation of intensity-weighed
effective opacity factors, which account for the presence of tens of
thousands of spectral lines, dominated by Fe and Ni. Based on 
{\sc isa}-wind atmosphere
calculations, the Monte Carlo (MC) code determines the line blanketing
factors. An iterative procedure is used, such that blanketing factors 
are used by the non-LTE code to calculate a new atmosphere,  which
in turn is used to calculate new blanketing factors. A few iterations 
are generally sufficient. This is due to the fact that the scattering 
and absorption factors are not very sensitive to the specific model 
parameters.

The MC method deals with the radiative transfer in the correct manner, 
except that the ionization and excitation equilibrium of metal species 
is approximate.
This dictates which lines are efficient at capturing photons for each 
point in the atmosphere. The ionization structure of H and He in the
MC code is derived from {\sc isa}-wind. For other species, an 
representative ionization temperature is used, obtained from the 
principal ionization stages for each metal that is considered in 
{\sc isa}-wind. The MC code calculated ionization stratification 
can therefore be compared with that of {\sc isa}-wind for consistency.

Since line-line interactions are not accounted for in the MC calculation,
explicit correction factors, named `photon loss' factors, need to 
be considered to allow for interaction between important resonance lines 
(He\,{\sc ii} $\lambda$303, He\,{\sc i} $\lambda$584) and 
nearby metal lines (Schmutz 1997). We have calculated photon loss factors 
from the opacities of He\,{\sc ii} $\lambda$303 and lines at nearby 
wavelengths, and find that line-line interactions are negligible for the 
stellar temperatures of WNL stars.  (Pasquali et al. (1997) previously 
adopted a factor of 10$^{-4}$ at He\,{\sc ii} $\lambda$303 for WNL stars.) 
Additionally, we have determined that line-line interaction between 
He\,{\sc i} $\lambda$584 and adjacent metal lines would not affect the
ionization balance or derived stellar parameters for WR124.

\section{Stellar analysis}

Unblanketed non-LTE spectroscopic
analyses of WR124 by Hamann et al. (1993) and Crowther et al. (1995a) revealed
$T_{\ast} \approx$30--34kK, $\log(L/L_{\odot})$$\sim$5.3--5.4,
$\mdot$$\sim$10$^{-4.2}$$M_{\odot}$yr$^{-1}$, and
$v_{\infty}$$\sim$700 km\,s$^{-1}$. The latter study included hydrogen,
helium, carbon and nitrogen, deriving H/He$\sim$0.6 by number.
Hamann \& Koesterke (1998a) assumed this hydrogen content, and 
derived equivalent stellar parameters based on an analysis of 
the nitrogen spectrum. In contrast, Esteban et al. 
(1993) used properties of M1--67 to conclude that 
$T_{\ast}$$\le$30kK, although 
a suitable unblanketed model was not found from the grid 
of Schmutz et al. (1992). The best agreement was obtained with the
only line blanketed models available at that time, namely
those for the WN9 star R84=Brey~18 ($T_{\ast}$=28.5kK) from 
Schmutz et al. (1991). A spectral comparison of WN8 
and WN9 stars is presented by Crowther \& Smith (1997).

\subsection{Analysis technique}

Our approach follows that of Crowther et al. (1995a), such that diagnostic
optical lines of He\,{\sc i} ($\lambda$5876), He\,{\sc ii} ($\lambda$4686)
and H\,{\sc i} (H$\beta$+He\,{\sc ii} $\lambda$4859) are chosen to derive
the stellar temperature, mass-loss rate, luminosity and hydrogen content.
In the absence of high resolution UV observations, a wind velocity of
710~km\,s$^{-1}$ is obtained from optical He\,{\sc i} P Cygni line profiles.
A $\beta$=1 velocity law is adopted, since this reproduces the optical, near-IR
and mid-IR spectra of a similar WN8 star, WR147 (Morris et al. 1999).

\subsection{Results}

In Fig.~\ref{fig2}, we compare spectroscopic observations of WR124
with the synthetic spectrum obtained from {\sc cmfgen}, allowing for both
line blanketing and clumping. The level of agreement 
between optical observations and
theory in Fig.~\ref{fig2} is excellent, except that the He\,{\sc i}
absorption components are often predicted to be too strong.

We also include comparisons with H and K band 
UKIRT observations of WR124 in Fig.~\ref{fig2}. Again, excellent
agreement is reached. Note that UKIRT observations of 
He\,{\sc ii} 3.09$\mu$m are also well reproduced. The case 
of He\,{\sc i} 2.058$\mu$m deserves special comment. 
Crowther et al. (1995a) discussed the difficulties in
predicting emission at He\,{\sc i} 2.058$\mu$m for the WN8 star 
HD\,86161, in that the predicted emission was far too weak.
The strength of 2.058$\mu$m (and He\,{\sc i} $\lambda$5016) is
very sensitive to the optical depth of the He\,{\sc i} 
$\lambda$584 transition. Crowther et al. (1995a) attributed
this failure to the neglect of line blanketing in the extreme-UV.
Our results support this claim for the case of WR124, the predicted
2.058$\mu$m emission now is in excellent agreement with observation.

Crowther et al. (1995a) obtained a similar level of consistency 
in their study, based on an identical optical
data set, except that electron scattering wings were strongly 
overestimated because of their assumption of homogeneity (their Fig.~10).
Allowing for clumping takes away this discrepancy, such
that electron scattering wings are well reproduced with an assumed
volume filling factor of $f$=10\%, with a corresponding decrease in the
required mass-loss rate by a factor of $\sim$3. As discussed by Hillier \&
Miller (1999) and Hamann \& Koesterke (1998b), a unique determination of the
filling factor is not possible, such that an uncertainty of $\pm$50\%
remains on the precise mass-loss rate. 

The stellar parameters for this model are
compared with those from the unblanketed, homogeneous analysis of
Crowther et al. (1995a) in Table~\ref{table3}. We find that the inclusion
of line blanketing has only a minor effect on the derived parameters
for WNL stars. Nevertheless, also considering the effects of clumping
on the mass-loss rate, the wind performance factor,
$\mdot v_{\infty}/(L_{\ast}/c)$, is reduced from 10 to $\sim$2!

\begin{figure}
\epsfxsize=17cm \epsfbox[90 400 610 570]{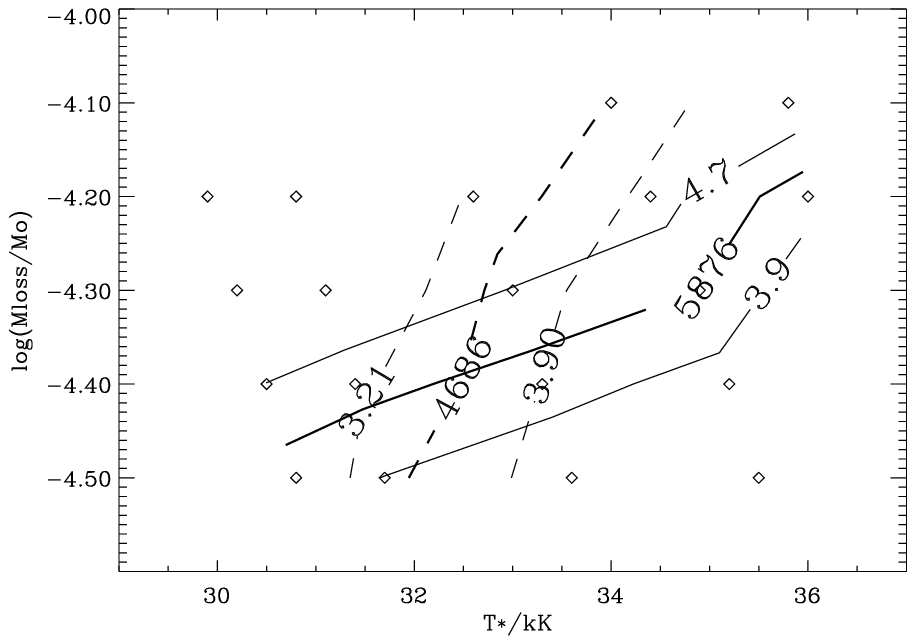}
\caption{Contour lines of equal rectified line peak intensity for 
He\,{\sc i} $\lambda$5876 (4.3, solid)  and He\,{\sc ii} $\lambda$4686 
(3.55, dashed), in the stellar $T_{\ast}$--log $\mdot$ plane. The thick 
contours represent the observed values, while the thin contours 
show the $\pm$10\% error margin. The rhomboids mark the parameters 
for which models were calculated.}
\label{fig2b}
\end{figure}

Turning to {\sc isa}-wind, we present a contour plot for the dependence
of our diagnostic He\,{\sc i--ii} lines on temperature and mass-loss for WR124.
We prefer to use a comparison in terms of peak intensity rather than line
equivalent width since the former approach was followed for {\sc cmfgen}.
Our figure allows us to quantify uncertainties in parameters, namely 
$\pm$1000K in temperature and $\pm$0.1 dex in mass-loss rate. Since
these represent formal uncertainties with {\sc isa}-wind, how do
the derived stellar parameters compare with results from {\sc cmfgen}?
Table~\ref{table3} reveals that good agreement is obtained
between the two codes in derived stellar parameters, including the 
hydrogen content. This comparison demonstrates the validity of the 
Sobolev approximation for WR analyses, and represents an 
important result from the present study.

In Fig.~\ref{fig2a}, we compare identical optical observations of WR124
with the synthetic spectrum obtained from the {\sc isa-}wind contour plot, 
allowing for  line blanketing. The quality of line fits 
is comparable to {\sc cmfgen},
except that (i) the region around $\lambda$4100 is more severely 
underestimated, due to the neglect of Si\,{\sc iv} $\lambda$4088-4116,
(ii) He\,{\sc i} $\lambda$5015 is too weak; (iii) He\,{\sc ii} $\lambda$4200,
$\lambda$4542 and $\lambda$5412 are in better agreement than {\sc cmfgen}. 
In  the near-IR, He\,{\sc i} 2.058$\mu$m 
emission is predicted to be somewhat too strong, as is
He\,{\sc i} 2.112$\mu$m emission, with Br$\gamma$ too weak. 

\begin{table}
\begin{center}
\caption{A comparison of stellar parameters obtained for WR124
using {\sc cmfgen} (unblanketed and blanketed) and {\sc isa}-wind (blanketed).
The unblanketed results are taken from Crowther et al. (1995b),
based on identical observations. In all cases, a standard velocity law 
with $\beta$=1 is assumed.}
\label{table3}
\begin{tabular}{cccc}
\hline
Parameters & \multicolumn{2}{c}{{\sc cmfgen}} & {\sc isa}-wind \\
Filling factor, $f$ & 100\% & 10\% &  100\% \\
Blanketed?  & no      & yes   &  yes \\
\hline                                   
$T_{\ast}$ (kK)                 & 33.5   & 32.7 &  32.8    \\
$R_{\ast}$ ($R_{\odot}$)        & 17.1   & 18.0 & 16.2 \\ 
log($L_{\ast}/L_{\odot}$)      & 5.52    &  5.53&  5.44    \\ %
log($\mdot/\sqrt{f}$) $M_{\odot}$yr$^{-1}$&$-$4.1 &$-$4.2& $-$4.4     \\
$v_{\infty}$ (km\,s$^{-1}$)    &   710   & 710   & 710 \\
H/He                           & 0.6     & 0.7 & 0.8     \\
N/He                           & 0.004   & 0.006 & 0.003     \\
log $Q_{0}$ (s$^{-1}$)         & 49.05   & 49.05& 49.00 \\
log $Q_{1}$ (s$^{-1}$)         & 48.19   & 43.88& 47.91 \\
\hline
\end{tabular}
\end{center}
\end{table}

\begin{figure}
\epsfxsize=8.4cm \epsfbox[90 80 465 710]{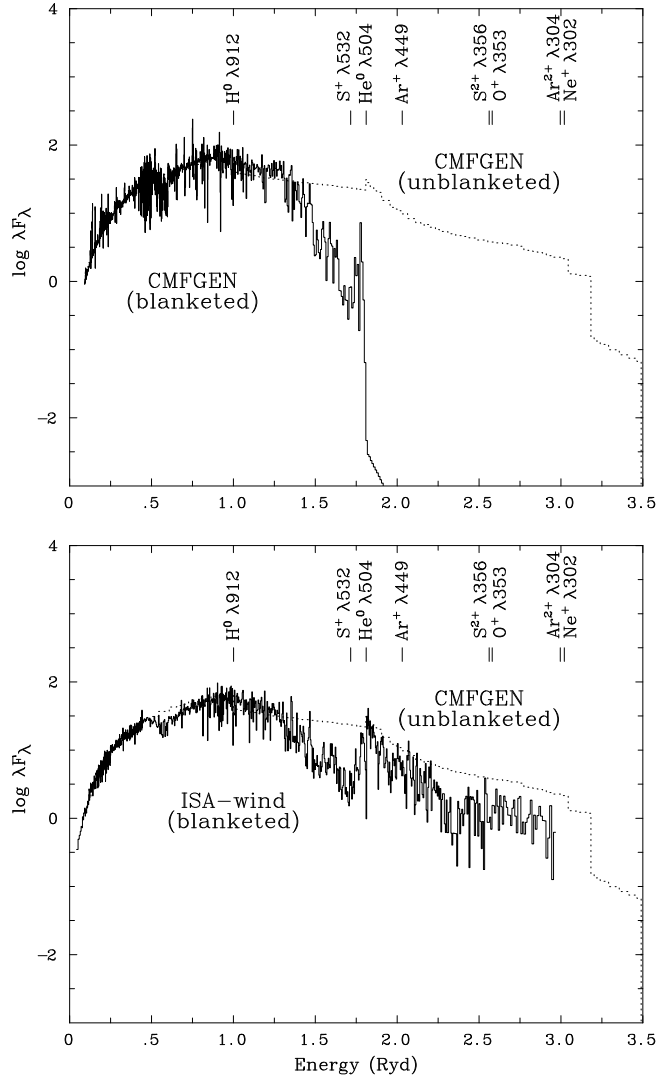}
\caption{(upper panel:) Comparison between the Lyman ionizing distribution of
unblanketed {\sc cmfgen} (dotted lines) and blanketed {\sc cmfgen} 
(solid lines) for WR124, revealing large differences; (Lower panel)
as above, except for unblanketed {\sc cmfgen} (dotted lines) and
blanketed {\sc isa}-wind (solid lines). Note that these diagrams are
designed to emphasise the extreme UV spectral region, important for
the nebular ionization, and not the overall energy distribution.}
\label{fig3}
\end{figure}

\subsection{Comparison of ionizing flux distributions}

Our three solutions indicate very similar stellar properties 
for WR124 and are equally successful at reproducing its optical
spectrum. How do the Lyman ionizing flux distributions
of these models compare? 

The unblanketed {\sc cmfgen} ionizing
energy distribution is shown in Fig.~\ref{fig3}. It is
relatively flat, with a small jump beyond the 
He\,{\sc i} edge at 1.81\,Ryd (24.6 eV or 504\AA), and a non-negligible
flux up to 3.2\,Ryd (286\AA). Fig.~\ref{fig3} reveals that the
blanketed {\sc cmfgen} distribution is much softer than
the unblanketed solution, showing a large deficit for energies
$\ge$1.3Ryd (i.e. $\lambda\le$700\AA), such that a very low flux is 
emitted at energies beyond the He\,{\sc i} edge. In addition, 
this distribution  
shows a moderate excess between the Lyman edge and 1.25\,Ryd relative
to the unblanketed case, which explains why its bolometric luminosity is
essentially identical. 

{\sc isa-}wind shows an intermediate energy 
distribution, with a depression around 1.7\,Ryd (540\AA), but a 
relatively hard flux 
beyond the He\,{\sc i} edge, up to energies of 2.9\,Ryd (310\AA). 
Possible explanations for the differences in ionizing fluxes between
the two blanketed cases are discussed in Sect.~\ref{differences}.

\section{Photo-ionization modelling}

{}From the usual
diagnostic diagram relating H$\alpha$/[S\,{\sc ii}] to [S\,{\sc ii}]
$\lambda$6717/$\lambda$6731 (see Sabaddin et al. 1977) we
find that M1--67 falls in the photo-ionization dominated region.
Therefore we use the general purpose photo-ionization code
{\sc cloudy} (Ferland 1996) for our analysis. The ionising radiation fields
presented in the previous section are expected to yield different predicted
nebular properties, when input into {\sc  cloudy}, which can be
quantitatively compared with spectroscopic nebular observations
taken from  Esteban et al. (1991).

\subsection{Description of the calculations}

We constructed photo-ionization models using {\sc cloudy} (v90.04) as
described in Ferland (1996) and Ferland et al. (1998).
Comparisons with other photo-ionization codes are provided by
Ferland et al. (1995). The nebula are represented by a sphere
of variable gas density, $n$, and filling factor, $\epsilon$, with
a central cavity that is ionized and heated solely by the
UV radiation of a single central star.  Nebular fluxes are predicted,
given input abundances, flux distributions and physical parameters.
We use a volume filling factor of $\epsilon$=0.05 as estimated by Solf \& Carsenty (1982).
The
de-projected radial density profile obtained by
Grosdidier et al. (1998) was used, together with an outer (inner) nebula
boundary at 50$''$ (8$''$), which corresponds to 1.25 pc (0.2 pc) using
a distance of 5 kpc. 

Grosdidier et al. (1998) provided a
measurement of the integrated nebular H$\alpha$ flux, after correction
for [N\,{\sc ii}] contamination. We obtain an integrated de-reddened 
H$\alpha$ flux using the reddening of $E_{\rm B-V}$=1.3 mag, determined in 
Sect.~\ref{reddening}, namely 
$I$({H$\alpha$)=5.45$\times$10$^{-10}$ erg\,cm$^{-2}$s$^{-1}$.
In addition, long slit nebular line intensities of Esteban et al. (1991)
(their position A) are de-reddened according to this value of $E_{\rm B-V}$.
Although the majority of previous stellar and nebular reddening determinations
are in reasonable agreement with our value, Esteban et al. 
(1991) obtained a substantially lower value of $E_{\rm B-V}$=0.90 mag for 
M1--67 using H$\alpha$/H$\beta$. Consequently, the 
H$\alpha$/H$\beta$ ratio, de-reddened according to our stellar 
reddening of $E_{\rm B-V}$=1.3 is far from the usual Case B value.

In order to assess the impact of alternative reddening determinations on
the nebular properties, we have also calculated the integrated 
H$\alpha$ flux, following the reddening determination of 
Esteban et al. (1991), i.e. $E_{\rm B-V}$=0.9, so that 
$I$(H$\alpha)$=2.35$\times$10$^{-10}$ erg\,cm$^{-2}$s$^{-1}$.
For this case, long slit nebular line intensities are taken directly
from Esteban et al. (1991). Note that identical stellar ionizing flux
distributions to those obtained for the higher reddening case are used 
in this analysis, except that the absolute K-band magnitude is adjusted
by 0.15\,mag to $M_{\rm K}=-$6.1\,mag, for consistency with the adopted
distance of 5\,kpc. Results from this alternative approach are very similar
to those discussed here for $E_{\rm B-V}$=1.3 mag.

The electron density at the inner boundary was selected so that the predicted
H$\alpha$ flux agreed with observations. For the $E_{\rm B-V}$=0.9 mag case, 
typical quantities are $N_{e}$=1,350 cm$^{-3}$ at 
the inner boundary (8$''$ or 0.2 pc), and 320 cm$^{-3}$ at the outer boundary 
(50$''$ or 1.25pc). Similar values were measured by Esteban et al. 
(1991), namely $\sim$1,000 cm$^{-3}$ at a distance of 10--25$''$ and 
$\sim$200 cm$^{-3}$ for $\ge$30$''$ (their region W).

Abundances are assumed to be typical of Galactic H\,{\sc ii} regions, except
for N, O and S, which are adapted from Esteban et al. (1991). For each
reddening, electron densities and temperatures were obtained from 
the de-reddened [S\,{\sc ii}] $\lambda$6716/$\lambda$6731 and [N\,{\sc ii}]
$\lambda$5755/$\lambda$6584 ratios using the {\sc ratio} program
(Adams \& Howarth, priv. comm.). For $E_{\rm B-V}$=0.9 we obtained
$N_{e}$=10$^{2.9}$ cm$^{-3}$ and $T_{e}$=6,000K, while 
$N_{e}$=10$^{3.0}$ cm$^{-3}$ and $T_{e}$=6,200K for $E_{\rm B-V}$=1.3.
Subsequently, ionic abundances were obtained by solving the equations of 
statistical equilibrium using the {\sc equib} program
(Adams \& Howarth, priv. comm.), and are given in Table~\ref{table4}
for each reddening. The usual elemental diagnostics were chosen, 
[N\,{\sc ii}] $\lambda$6584, 
[O\,{\sc ii}] $\lambda$3727, [S\,{\sc ii}] $\lambda$6731 and [S\,{\sc iii}]
$\lambda$9069, with no ionization correction factors for nitrogen
(since O$^{2+} \ll$ O$^{+}$). Also, we have adopted He/H$\sim$0.2, since 
helium is expected 
to be moderately enriched. Solar abundance ISM dust grains 
were assumed, although the precise choice is not critical.

\begin{table}
\caption[]{Comparison between observed and predicted 
nebular properties of M1-67 from {\sc cloudy}, based on various ionizing 
flux distributions. Nebular line measurements are taken from slit position 
A of Esteban et al. (1991), de-reddened according to $E_{\rm
B-V}$=1.3 (Sect.~\ref{reddening}) and given relative to H$\beta$=100, including 
{\it formal} errors from Esteban et al., neglecting systematic (calibration) 
uncertainties. I(H$\beta$) represents the de-reddened, integrated H$\beta$ flux 
received at the Earth (erg\,cm$^{-2}$s$^{-1}$), obtained 
via the H$\alpha$ flux measured by Grosdidier et al. (1998).}
\label{table4}
\begin{tabular}{l@{\hspace{5mm}}c@{\hspace{4mm}}c@{\hspace{4mm}}c@{\hspace{4mm}}c}
\hline
Quantities             & Observed &{\sc cmfgen}&{\sc cmfgen}&{\sc isa}-wind\\
Blank?                  &        &no    & yes  & yes\\  
\hline
log I(H$\beta$)       &$-$9.7 &$-$9.7 &$-$9.7 &$-$9.7 \\  
$\epsilon$            &       & 0.05  & 0.05  & 0.05  \\
$N_{e}(r=$0.2pc)      &       & 2050  & 1650  & 2000  \\  
$N_{e}(r$=1.25pc)     &       &  480  &  380  &  460  \\  
12+log N/H            & 8.21  & 8.21  & 8.21  & 8.21  \\
12+log O/H            &8.09  & 8.09  & 8.09  & 8.09  \\
12+log S/H            &6.64  & 6.64  & 6.64  & 6.64  \\
He/H                  & 0.2  & 0.2   & 0.2   & 0.2   \\
\\
3727 {\rm [O\,{\sc ii}]}   &21$\pm$2    &37     &52    &52     \\ 
5007 {\rm [O\,{\sc iii}]}  &$\le$1:     &316    &0.0   &148    \\ 
5755 {\rm [N\,{\sc ii}]}   &0.7$\pm$0.1 &1.1    &1.9   &1.3    \\ 
5876 He\,{\sc i}             &1.6$\pm$0.3 &30     &0.0   &31     \\ 
6548 {\rm [N\,{\sc ii}]}   & 71$\pm$1   &24     &113   &45     \\ 
6563 H$\alpha$               &194$\pm$1$^{\ast}$&284&292 &287    \\
6584 {\rm [N\,{\sc ii}]}   &204$\pm$2   &71     &333   &134    \\ 
6717 {\rm [S\,{\sc ii}]}   & 8.4$\pm$0.2&1.8    &12.5  &2.7    \\ 
6731 {\rm [S\,{\sc ii}]}   &10.3$\pm$0.2&1.9    &12.8  &2.8    \\ 
7135 {\rm [Ar\,{\sc iii}]} &0.2:        &68     &0.1   &50     \\ 
9069 {\rm [S\,{\sc iii}]}  & 5.5$\pm$0.3&30     &9.5   &24     \\ 
9546 P8                      &1.9$\pm$0.2 &3.8    &3.9   &3.9    \\ 
\\                          
N$^{2+}$/N$^+$               &            &10     &0.00   &3     \\ 
O$^{2+}$/O$^+$               & $\le$0.02  &5.5    &0.00   &1.5   \\ 
S$^{2+}$/S$^+$               & 0.9        &35     &1.2    &17    \\ 
Ar$^{2+}$/Ar$^+$             &            &57     &0.00   &25    \\ 
T$_e$(K)                     &6,200       &9,800  &6,900  &8,500 \\ 
\hline
\end{tabular}
\newline
$\ast$: See text for an explanation why the H$\alpha$ flux
does not agree with the Case~B value using our stellar reddening.
\end{table}

\subsection{Results from photo-ionization modelling}

Esteban et al. (1993) identified a large discrepancy between the nebular 
fluxes of M1--67 as determined by unblanketed model atmospheres
(Schmutz et al. 1992) and that observed. In particular, they were unable
to reproduce the observed $T_{e}$ and S$^{+}/$S$^{2+}$ ratio. 
Closest agreement was achieved for the line blanketed model 
of R84 (WN9) from Schmutz et al. (1991). We now investigate whether this 
discrepancy remains, based on improved stellar models, including 
those for which line blanketing is incorporated.
Comparisons between predicted and observed nebular properties 
are presented  in Table~\ref{table4}, including results from
both the reddening of Esteban et al. (1991), or the present value,
to illustrate the possible range of line strengths and plasma conditions.
Lamers et al. (1999) discuss various aspects of the reliability of 
generally derived nebular conditions and abundances, of relevance to 
the present comparisons.

\subsubsection{CMFGEN}

From Table~\ref{table4}, the discrepancy between the observed plasma
conditions and photo-ionization modelling from the unblanketed {\sc cmfgen}
analysis is more acute than Esteban et al. (1993) obtained, for
both reddening solutions. (A higher stellar temperature is imposed here by 
modelling the stellar spectrum, including the allowance for light metals, 
whereas Esteban et al. did not adopt any stellar constraint).

In this case, the ionization balance in the nebula is much higher than 
observed, with S$^{2+}/$S$^{+}$ overestimated by a factor of 40. 
In addition, the mean $T_{e}$ is far too high, with the predicted
range from 11,200K at the inner nebula to 9,500K at the outer boundary,
contrasting sharply with the observed value of 6,200K. 
Recalling Fig.~\ref{fig3}, a relatively strong flux at energies greater
than the S$^{2+}$ edge at 356\AA\ indicates that the predicted nebular ionization 
balance implies S$^{2+} \gg $ S$^{3+} > $ S$^{+}$. Similarly for
oxygen, the strong flux predicted below the O$^{+}$ edge at 353\AA\ 
implies an error in the O$^{2+}/$O$^{+}$ ratio of order $\ge$300!
Consequently, [O\,{\sc iii}] $\lambda$5007 is predicted to have an intensity
that is greater than H$\beta$, yet [O\,{\sc iii}] $\lambda$5007 is barely 
detected in M1--67. Similar problems relate to 
[N\,{\sc ii}] $\lambda$6584, He\,{\sc i} $\lambda$5876 and
[Ar\,{\sc iii}] $\lambda$7135.

In contrast, the line blanketed {\sc cmfgen} model,
with a much softer extreme UV energy distribution, produces
a much lower predicted electron temperature, in the range 6,600--7,900K,
now in reasonable agreement with observations. From Table~\ref{table4},
almost every nebular line is predicted to be within a factor of two 
of the de-reddened value, with the predicted ionization balance in 
good agreement with observations.  

An explanation
for the success of this model can be understood by recalling Fig.~\ref{fig3} 
in which negligible flux is predicted at energies greater than the 
He$^{0}$ edge at 504\AA, and indeed only a small flux beyond the 
S$^{+}$ $\lambda$532 edge. Consequently, [S\,{\sc ii}] $\lambda$6731 and 
especially 
[S\,{\sc iii}] $\lambda$9069 are fairly well matched for either case,
although the predicted ionization balance is actually marginally 
too low: S$^{+} \simeq $ S$^{2+} \gg $ S$^{3+}$. Another important
success of this model
is that negligible flux is predicted above the O$^{+}$ 353\AA\ edge,
so that the observed absence of [O\,{\sc iii}] $\lambda$5007 is accounted for.

However, the ionization of the model predicts negligible 
He\,{\sc i} $\lambda$5876 and [Ar\,{\sc iii}] $\lambda$7135 emission,
yet these are both observed, albeit weakly. Note that the precise choice 
of the He/H abundance ratio in the nebula does not play a major role 
in the predicted He\,{\sc i} $\lambda$5876 strength, since it is primarily 
determined by the number of He\,{\sc i} ionizing photons. Less crucial 
deficiencies  of this model are that [N\,{\sc ii}] $\lambda$6584 and 
[O\,{\sc ii}] $\lambda$3727 are predicted to be too strong, although 
[N\,{\sc ii}]/H$\alpha$ (independent of reddening uncertainties) is 
in good agreement.

\subsubsection{ISA-wind}

Turning to {\sc isa}-wind, its energy distribution lies intermediate
between the unblanketed and blanketed {\sc cmfgen} models, such that 
the nebular ionization structure is predicted to be somewhat higher than the 
observed properties, with $T_{e}$ overestimated by 40\%.
Once again, a significant ionizing flux is predicted to lie at energies
above the O$^{+}$ edge, so that the predicted O$^{2+}/$O$^{+}$ 
ratio is a factor of $\ge$75 times too high, with very strong [O\,{\sc iii}] 
$\lambda$5007 predicted. Similarly, the flux above the S$^{2+}$ and He$^{0}$
edges are too strong, so that that the S$^{2+}/$S$^{+}$ ratio is 
in error by a factor of 20, and He\,{\sc i} $\lambda$5876 is strongly 
overestimated. 

\subsection{Influence of uncertain nebular properties}

How dependent are the predicted nebular properties on our assumed 
filling factor, $\epsilon$, or geometry?  As a test, we have carried out 
calculations for the blanketed {\sc cmfgen} case with
$E_{\rm B-V}$=1.3 mag.

For $\epsilon$=0.025 (a factor of two smaller), the required electron 
density is $\sqrt{2}$ higher, while predicted nebular properties and
line strengths properties are barely affected, with the
electron temperature unchanged. The greatest influence of a decrease in
filling factor is a decrease in the sulphur ionization balance, from 
S$^{2+}$/S$^{+}$=1.2 to 0.9, with changes in nebular line strengths
of up to 20\% for [S\,{\sc ii-iii}]. Comparable increases in 
S$^{2+}$/S$^{+}$ and decreases in electron density are obtained 
for higher filling factors. 

Similarly, we have compared results with a constant, rather than exponentially 
decreasing, density throughout the nebula. For $\epsilon$=0.05, the required 
electron density is $N_{e}$=530 cm$^{-3}$. Again, the influence is generally
small, with the electron temperature unchanged and line strengths modified
by up to 10\%. Once again, the effect on sulphur is greatest, 
such that the ionization balance is decreased from S$^{2+}$/S$^{+}$=1.2 
to 1.0 and changes in [S\,{\sc ii-iii}] strengths of up to 20\%.

From these calculations, we consider that the nebular predictions are
robust, for the blanketed {\sc cmfgen} case at least, except that the
sulphur line strengths and ionization balance suffer from uncertainties
of order $\pm$50\%. Equivalent tests for the unblanketed {\sc cmfgen} case
indicate similar uncertainties for the ionization balance of oxygen and
nitrogen, although the effect on emission lines from [O\,{\sc ii--iii}]
is minor.

Finally, one further note of caution is necessary. 
Lamers et al. (1999) emphasise that nebular line fluxes and
(absolute) abundances may be sensitive to clumping, of particular concern 
for the M1--67 nebula. Although beyond the scope of the present
work, one would ideally like to carry out a re-evaluation of the nebular 
abundances and conditions of M1-67, accounting for its highly clumped 
nature.

\subsection{Why are predicted ionizing fluxes different?}\label{differences}

A definitive explanation for the difference in ionizing spectra produced
by the blanketed {\sc cmfgen} and {\sc isa}-wind models 
shortward of $\lambda$700\AA\ is not straightforward, given the
differences in their treatment of the radiative transfer, blanketing, 
atomic models  and atomic data. Nevertheless, several differences in 
the alternative techniques deserve consideration.

The most promising reason for the differences is that {\sc cmfgen} 
allows for photons that are absorbed in lines and re-emitted at other 
(longer) wavelengths. In contrast, no such channel is available in 
the {\sc isa}-wind MC calculations, thereby overestimating the ionizing flux.
Other possible explanations, such as photon loss operating 
in {\sc cmfgen} at the He\,{\sc i} $\lambda$584 resonance line, or an 
incorrect iron ionization balance being used for the MC 
calculation in {\sc isa}-wind, appear to be excluded (Sect.~\ref{ISA}).

In addition, the number of lines treated in both cases is incomplete, 
so that blocking will be underestimated, and the predicted ionizing fluxes
overestimated. Recall that the number of lines considered by {\sc cmfgen}
is relatively small, principally treating Fe\,{\sc iii-v}, while
{\sc isa}-wind considers many more lines and ionization stages, although 
this is clearly not complete (solely lines between measured energy levels 
are included). Depending on the strength of this effect, softer ionizing 
energy distributions would result, leading to improved agreement with
nebular properties for {\sc isa}-wind. 

In spite of the modelling and nebular difficulties, we conclude that
the {\sc cmfgen} line blanketed ionizing energy distribution for WR124 
provides excellent consistency with the nebular properties of M1--67, 
and solves the problem that was identified by Esteban et al. (1993) for 
this system based on unblanketed models. Use 
of {\sc isa-}wind leads to consistent stellar properties for WR124,
although the consistency with observed plasma conditions for M1--67 are 
somewhat poorer than for {\sc cmfgen}. 

\section{Conclusions}

We have used WR124 and its associated nebula  M1--67 to assess the
reliability of Lyman ionizing flux distributions from current stellar
models. For WR124, we find that stellar properties obtained with two
line blanketed, non-LTE extended model atmospheres, 
{\sc isa}-wind (de Koter et al. 1997) and {\sc cmfgen} 
(Hillier \& Miller 1998) are in excellent agreement. However, 
comparisons with the observed properties of M1--67 favour the ionizing
energy distribution obtained by {\sc cmfgen} rather than {\sc isa}-wind, 
reconciling previous inconsistencies identified by Esteban et al. (1993).
Certainly, more general  conclusions  await comparisons for a variety 
of stars.

In the future we propose to extend our study to WR stars of different
spectral type. One well known example is HD\,192163 (WR136) and its
ring nebula NGC\,6888, for which Esteban et al. (1993) also identified
a major inconsistency. However, in contrast with M1--67, there is evidence 
that nebular emission in NGC\,6888 may be produced by shocks rather 
than photo-ionization. Further, the properties of other Galactic WR 
nebulae are often poorly known, such that different stellar ionizing 
distributions can reproduce the  nebular properties by selecting, 
for example, alternative electron densities 
or shell thickness. WR nebulae in the LMC may provide more suitable 
candidates (Dopita et al. 1994).

In general, study of young stellar populations in external galaxies
via associated H\,{\sc ii} regions relies heavily on suitable stellar
ionizing flux distributions (e.g. Leitherer et al. 1999). Currently,
the only WR models that are generally available at this time are 
unblanketed, pure helium energy distributions (Schmutz et al. 1992). 
The calculation  of a large multi-parameter grid of line-blanketed models 
remains a formidable computational challenge, unless codes such as 
{\sc isa}-wind can be relied upon. Crowther (1999) finds that, in 
general, the H\,{\sc i} and He\,{\sc i} ionizing fluxes of early-type
WR stars are not greatly affected by line blanketing. However, the additional 
blanketing from C and O in early WC stars produces a softer ionizing spectrum 
than for early WN stars of identical temperatures, with negligible flux 
emitted  $\lambda \le$300~\AA. Indeed, unblanketed models are probably 
to be preferred for the study of young starburst regions in very 
metal deficient regions. However, the (uncertain) role that metallicity may 
play on mass-loss rates of Wolf-Rayet stars does need to be 
quantitatively investigated (e.g. Crowther et al. 1999).

\begin{acknowledgements}
We are especially grateful to 
Linda Smith and Luc Dessart for observing WR124 at UKIRT, 
and for useful discussions. PAC is grateful to the STScI, in 
particular Antonella Nota, for financial support from the Visitors Fund 
where this work was initiated, and ST-ECF for support on a 
subsequent visit.  PAC is funded by a Royal Society University 
Research Fellowship. OD acknowledges financial support from PPARC grant 
PPA/G/S/1997/00780. AdK acknowledges financial support from NWO
Pioneer grant 600-78-333 to R.B.F.M.~Waters and from NWO Spinoza grant 
08-0 to E.P.J.~van den Heuvel. The  U.K. Infrared Telescope is operated 
by the Joint Astronomy Centre on behalf of the Particle Physics and 
Astronomy Research Council. 
\end{acknowledgements}

\end{document}